\documentclass[prb,twocolumn,superscriptaddress,showpacs,preprintnumbers,amsmath,amssymb]{revtex4-1}

\usepackage{graphicx}
\usepackage{dcolumn}
\usepackage{bm}
\usepackage{color,ulem}

\begin{document}


\title{
Horizontal Line Nodes in Sr$_2$RuO$_4$ Proved by Spin Resonance
}

\author{Kazuki~Iida}\email{k_iida@cross.or.jp}
\affiliation{Department of Physics, University of Virginia, Charlottesville, Virginia 22904-4714, USA}
\affiliation{Neutron Science and Technology Center, Comprehensive Research Organization for Science and Society (CROSS), Tokai, Ibaraki 319-1106, Japan}

\author{Maiko~Kofu}
\affiliation{Department of Physics, University of Virginia, Charlottesville, Virginia 22904-4714, USA}
\affiliation{J-PARC Center, Japan Atomic Energy Agency (JAEA), Tokai, Ibaraki 319-1195, Japan}

\author{Katsuhiro~Suzuki}
\affiliation{Research Organization of Science and Technology, Ritsumeikan University, Kusatsu, Shiga 525-8577, Japan}

\author{Naoki~Murai}
\affiliation{J-PARC Center, Japan Atomic Energy Agency (JAEA), Tokai, Ibaraki 319-1195, Japan}

\author{Ryoichi~Kajimoto}
\affiliation{J-PARC Center, Japan Atomic Energy Agency (JAEA), Tokai, Ibaraki 319-1195, Japan}

\author{Seiko~Ohira-Kawamura}
\affiliation{J-PARC Center, Japan Atomic Energy Agency (JAEA), Tokai, Ibaraki 319-1195, Japan}

\author{Yasuhiro~Inamura}
\affiliation{J-PARC Center, Japan Atomic Energy Agency (JAEA), Tokai, Ibaraki 319-1195, Japan}

\author{Motoyuki~Ishikado}
\affiliation{Neutron Science and Technology Center, Comprehensive Research Organization for Science and Society (CROSS), Tokai, Ibaraki 319-1106, Japan}

\author{Shunsuke~Hasegawa}
\affiliation{Neutron Science Laboratory, Institute for Solid State Physics, University of Tokyo, Kashiwa, Chiba 277-8581, Japan}

\author{Takatsugu~Masuda}
\affiliation{Neutron Science Laboratory, Institute for Solid State Physics, University of Tokyo, Kashiwa, Chiba 277-8581, Japan}

\author{Yoshiyuki~Yoshida}
\affiliation{National Institute of Advanced Industrial Science and Technology, Tsukuba, Ibaraki 305-8565, Japan}

\author{Kazushige~Machida}
\affiliation{Department of Physics, Ritsumeikan University, Kusatsu, Shiga 525-8577, Japan}

\author{Seunghun~Lee}
\affiliation{Department of Physics, University of Virginia, Charlottesville, Virginia 22904-4714, USA}

\date{\today}

\begin{abstract}
We investigated the low-energy incommensurate (IC) magnetic fluctuations in Sr$_2$RuO$_4$ by the high-resolution inelastic neutron scattering measurements and random phase approximation (RPA) calculations. 
We observed a spin resonance with energy of $\hbar\omega_\text{res}=0.56$~meV centered at a characteristic wavevector $\mathbf{Q}_\text{res}=(0.3, 0.3, 0.5)$.
The resonance energy corresponds well to the superconducting gap $2\Delta=0.56$~meV estimated by the tunneling spectroscopy.
The spin resonance shows the $L$ modulation with a maximum at around $L = 0.5$.
The $L$ modulated intensity of the spin resonance and our RPA calculations indicate that the superconducting gaps regarding the quasi-one-dimensional $\alpha$ and $\beta$ sheets at the Fermi surfaces have the horizontal line nodes.
These results may set a strong constraint on the pairing symmetry of Sr$_2$RuO$_4$.
We also discuss the implications on possible superconducting order parameters.
\end{abstract}

\maketitle

Strontium ruthenate Sr$_2$RuO$_4$ with $T_\text{c}=1.5$~K~\cite{Sr2RuO4_Maeno_1994} has attracted a great deal of interest as a prime candidate for the chiral $p$-wave superconductor~\cite{pWave_1,pWave_2,Review_1,Review_2}; nuclear magnetic resonance (NMR)~\cite{Sr2RuO4_NMR_Ishida_1998} and polarized neutron scattering~\cite{Sr2RuO4_Neutron_Duffy_2000} measurements reported spin-triplet superconductivity whereas muon spin rotation~\cite{Sr2RuO4_muSR_Luke_1998} and Kerr effect~\cite{Sr2RuO4_Kerr_Xia_2006} measurements showed spontaneously time reversal symmetry breaking.
On the other hand, there are some experimental results such as the absence of the chiral edge currents~\cite{Sr2RuO4_SQUID_Kirtley_2007}, first-order superconducting transition~\cite{Sr2RuO4_HeatCapacity_Yonezawa_2013,Sr2RuO4_Magnetization_Kittaka_2014,Sr2RuO4_Theory_Amano,Sr2RuO4_Torque_Kikugawa_2016}, and strong $H_{c2}$ ($||$ $ab$) suppression~\cite{Hc2suppression}, all of which challenge the chiral $p$-wave superconductivity.
Recently, experimental and theoretical studies under an application of uniaxial pressure along $\langle100\rangle$ reported a factor of 2.3 enhancement of $T_\text{c}$ owing to the Lifshitz transition when the Fermi level passes through a van Hove singularity, raising the possibility of an even-parity spin-singlet order parameter in Sr$_2$RuO$_4$~\cite{Sr2RuO4_Pressure_Hicks_2014,Sr2RuO4_Pressure_Steppke_2017,Sr2RuO4_Pressure_Watson_2018,Sr2RuO4_Pressure_Barber_2018,Sr2RuO4_Pressure_Luo_2019,Sr2RuO4_Pressure_Sunko_2019,Sr2RuO4_Pressure_Li_2019}.
More recently, new NMR results demonstrated that the spin susceptibility substantially drops below $T_\text{c}$ provided that the pulse energy is smaller than a threshold, ruling out the chiral-$p$ spin-triplet superconductivity~\cite{Sr2RuO4_newNMR_Pustogow_2019,Sr2RuO4_newNMR_Ishida_2019}.
As such, these recent works turn the research on Sr$_2$RuO$_4$ towards a fascinating new era.

So far, various experimental techniques reported that the superconducting gaps of Sr$_2$RuO$_4$ have line nodes~\cite{Sr2RuO4_NMR_Ishida_2000,Sr2RuO4_PenetrationDepth_Bonalde_2000}, but the details of the line nodes, e.g. of the vertical or horizontal nature, are not uncovered yet.
The thermal conductivity measurements reported vertical line nodes on the superconducting gaps~\cite{Sr2RuO4_ThermalConductivity_Hassinger_2017}, whereas field-angle-dependent specific heat capacity measurements are indicating the horizontal line nodes~\cite{Sr2RuO4_HeatCapacity_Kittaka_2018,Sr2RuO4_Theory_Machida2}.
Since the complete information of the superconducting gaps can shed light on the symmetry of the pairing, exclusive determination of the direction of the line nodes in Sr$_2$RuO$_4$ is highly desirable.

The inelastic neutron scattering (INS) technique can directly measure the imaginary part of generalized spin susceptibility ($\chi''$) as a function of momentum ($\mathbf{Q}$) and energy ($\hbar\omega$) transfers, yielding information on the Fermi surface topology.
In addition, $\mathbf{Q}$ dependence of a spin resonance as a consequence of the Bardeen-Cooper-Schrieffer (BCS) coherence factor can provide information on the symmetry of the superconducting gap.
In Sr$_2$RuO$_4$, the most pronounced magnetic signal in the normal state is nearly two-dimensional incommensurate (IC) magnetic fluctuations at $\mathbf{Q}_\text{IC}=(0.3, 0.3, L)$ owing to the Fermi surface nesting between (or within) the quasi-one-dimensional $\alpha$ and $\beta$ sheets consisted of the $d_{zx}$ and $d_{yz}$ orbitals of Ru$^{4+}$~\cite{Sr2RuO4_Theory_Mazin_1999,Sr2RuO4_Neutron_Sidis_1999,Sr2RuO4_Neutron_Iida_2011,Sr2RuO4_Neutron_Kunkemoller_2017}.
In contrast to the marked signal from the IC magnetic fluctuations, only weak signals due to ferromagnetic fluctuations originating from the two-dimensional $\gamma$ sheet ($d_{xy}$) are observed around the $\Gamma$ point~\cite{Sr2RuO4_Neutron_Braden_2002,Sr2RuO4_Neutron_Steffens_2018}.
Upon decreasing temperature below $T_\text{c}$, however, no sizable spin resonance regarding the IC magnetic fluctuations~\cite{Sr2RuO4_ResonanceTheory_Morr_2001} was observed at $\mathbf{Q}=(0.3, 0.3, 0)$ with energy close to the superconducting gap $2\Delta=0.56$~meV~\cite{Sr2RuO4_Neutron_Kunkemoller_2017} estimated by the tunneling spectroscopy~\cite{Sr2RuO4_Tunneling_Suderow_2009}.

Based on the horizontal line nodes model, the spin resonance is supposed to emerge at $\mathbf{Q}=(0.3, 0.3, L)$ with a finite $L$ but not at $(0.3, 0.3, 0)$ because of the sign change of the superconducting gap along $k_z$~\cite{Sr2RuO4_Theory_Machida2}.
The vertical line nodes model, on the other hand, expects the spin resonance at $\mathbf{Q}=(0.3, 0.3, 0)$ because of the sign change within the $k_z$ plane.
Therefore, the horizontal line nodes model can naturally explain the absence of the spin resonance at $\mathbf{Q}=(0.3, 0.3, 0)$, and search for the spin resonance at $(0.3, 0.3, L)$ is of particular interest.
In this letter, we investigate in detail the low-energy IC magnetic fluctuations in Sr$_2$RuO$_4$ focusing on their $L$ dependence using the high-resolution INS technique.
The experimental results, especially the $L$ dependence of neutron scattering intensities, are compared with the random phase approximation (RPA) calculations.

Three single crystals of Sr$_2$RuO$_4$ with a total mass of $\sim10$~g were grown by the floating-zone method~\cite{Sr2RuO4_Synthesis_Ikeda_2002,Sr2RuO4_Synthesis_Mao_2000}, and each crystal shows the superconducting onset temperature of $\sim1.35$~K.
They were co-aligned with the $(HHL)$ plane being horizontal to the scattering plane and attached to a closed-cycle $^3$He refrigerator.
We performed INS measurements at 0.3 and 1.8~K using the disk chopper time-of-flight (TOF) neutron spectrometer AMATERAS of the Materials and Life Science Experimental Facility (MLF) in Japan Proton Accelerator Research Complex (J-PARC)~\cite{AMATERAS}.
Frequency of the pulse shaping and velocity selecting disk choppers was 300~Hz, yielding the combinations of multiple incident neutron energies ($E_\text{i}$s) of 2.64, 5.93, and 23.7~meV with the energy resolutions of 0.046, 0.146, and 1.07~meV, respectively, at the elastic channel.
The software suite Utsusemi~\cite{Utsusemi} was used to visualize the TOF neutron scattering data.

\begin{figure}[t]
\includegraphics[width=8.4cm]{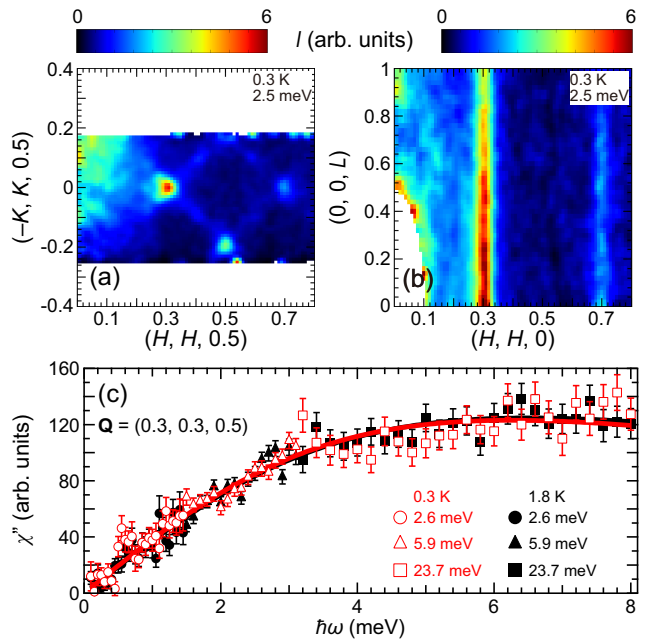}
\centering
\caption{
(Color online)
High-energy INS results of the IC magnetic fluctuations in Sr$_2$RuO$_4$.
(a, b) Constant-energy INS intensity maps in the (a) $(HK0.5)$ and (b) $(HHL)$ planes with the energy window of $[1.5, 3.5]$~meV at 0.3~K.
(c) $\chi''(\hbar\omega)$ spectra at $\mathbf{Q}_\text{IC}=(0.3, 0.3, 0.5)$ below and above $T_\text{c}$.
Data from different incident energies of neutrons $E_\text{i}=2.64$, 5.93, and 23.7~meV (circles, triangles, and squares) are combined for each temperature after background estimated by $\chi''(\hbar\omega)$ at $\mathbf{Q}=(0.525, 0.525, 0.5)$ is subtracted from each spectrum.
Solid lines are fitting results by the conventional relaxation response model.
}\label{Fig:HighE}
\end{figure}

Figure~\ref{Fig:HighE} summarizes the overall features of high-energy ($\hbar\omega\gg2\Delta$) IC magnetic fluctuations with $L = 0.5$ in Sr$_2$RuO$_4$.
Figure~\ref{Fig:HighE}(a) depicts the constant-energy INS intensity map in the $(HK0.5)$ plane with energy transfer of 2.5~meV at 0.3~K.
IC magnetic fluctuations are observed at $\mathbf{Q}_\text{IC}=(0.3, 0.3, 0.5)$, (0.7, 0.3, 0.5), and (0.7, 0.7, 0.5).
In addition to the IC magnetic fluctuations, the Fermi surface nesting also induces the ridge scattering connecting equivalent $\mathbf{Q}_\text{IC}$s around $(0.5, 0.5, 0.5)$.
To explore the energy evolution of the IC magnetic fluctuations with $L=0.5$, the INS intensity is converted to the imaginary part of the spin susceptibility $\chi''(\hbar\omega)$ via the fluctuation-dissipation theorem $\chi"(\hbar\omega)=(1-e^{-\hbar\omega/k_\text{B}T})I(\hbar\omega)$ after subtracting the background.
The $\chi''(\hbar\omega)$ spectra at $\mathbf{Q}_\text{IC}=(0.3, 0.3, 0.5)$ below and above $T_\text{c}$ are plotted in Fig.~\ref{Fig:HighE}(c).
The $\chi''(\hbar\omega)$ spectra above 1.0~meV are well fitted by the relaxation response model $\chi"(\hbar\omega)=\chi'\Gamma\hbar\omega/[(\hbar\omega)^2+\Gamma^2]$ where $\chi'$ is the static susceptibility and $\Gamma$ the relaxation rate [or the peak position of $\chi''(\hbar\omega)$], yielding $\Gamma=6.3(2)$~meV [6.5(2)~meV] at 0.3~K (1.8~K).
The observed IC magnetic fluctuations with $L=0.5$ are quantitatively consistent with the IC magnetic fluctuations with $L=0$ reported in the previous INS works~\cite{Sr2RuO4_Neutron_Sidis_1999,Sr2RuO4_Neutron_Iida_2011,Sr2RuO4_Neutron_Kunkemoller_2017,Sr2RuO4_Neutron_Braden_2002,Sr2RuO4_Neutron_Steffens_2018}.
This is reasonable since the IC magnetic fluctuations along $(0.3, 0.3, L)$ monotonically decrease in intensity with increasing $L$ [Fig.~\ref{Fig:HighE}(b)], representing the quasi-two-dimensional nature of the IC magnetic fluctuations in this energy region.
This is consistent with the quasi-one-dimensional band structures of the cylindrical $\alpha$ and $\beta$ sheets~\cite{Sr2RuO4_ARPES_Iwasawa_2010}.

\begin{figure*}[t]
\includegraphics[width=17.28cm]{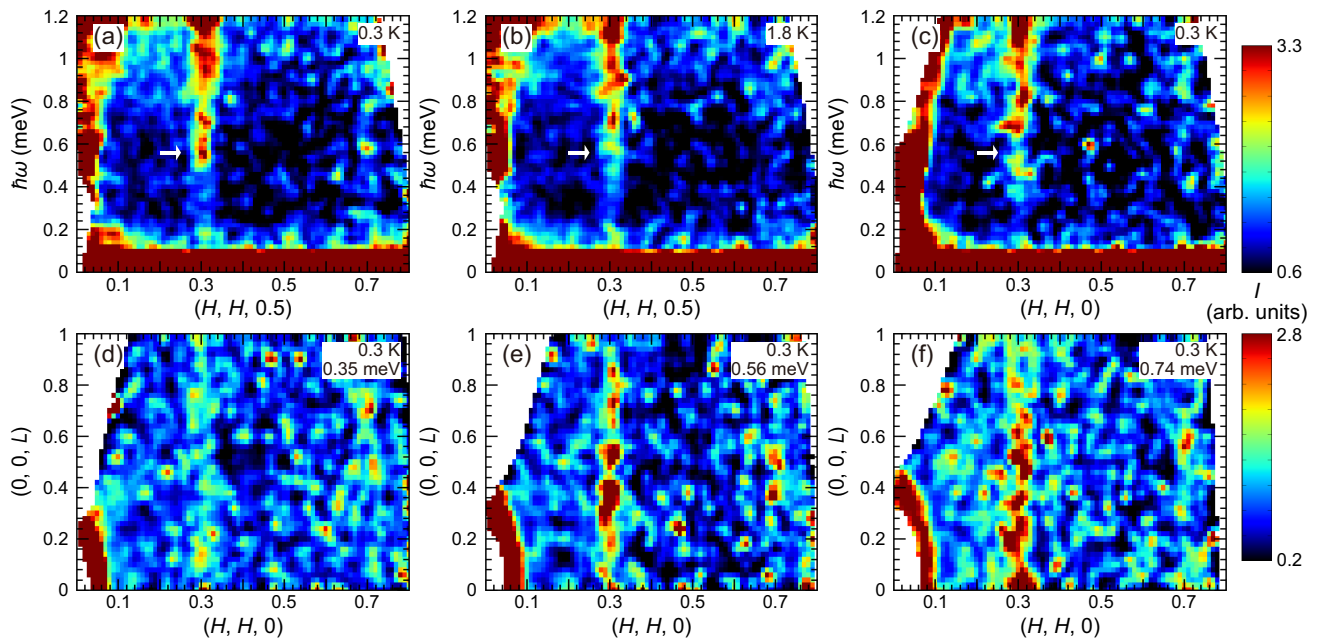}
\centering
\caption{
(Color online)
Low-energy INS intensity maps regarding the IC magnetic fluctuations in Sr$_2$RuO$_4$.
(a--c) Low-energy IC magnetic fluctuations (a) along $\mathbf{Q}=(H, H, 0.5)$ at 0.3~K, (b) along $(H, H, 0.5)$ at 1.8~K, and (c) along $(H, H, 0)$ at 0.3~K.
White arrows indicate the energy of the superconducting gap $2\Delta$ at $\mathbf{Q}_\text{IC}$.
(d--f) Constant-energy INS intensity maps in the $(HHL)$ plane at 0.3~K with the energy windows of (d) [0.23, 0.47]~meV, (e) [0.47, 0.65]~meV, and (f) [0.65, 0.83]~meV.
}\label{Fig:Map}
\end{figure*}

Let us turn to the low-energy ($\hbar\omega\simeq2\Delta$) IC magnetic fluctuations of Sr$_2$RuO$_4$.
We found that a spin resonance appears in the superconducting state at $\mathbf{Q}_\text{res}=(0.3, 0.3, 0.5)$ and $\hbar\omega_\text{res}$ = $2\Delta$ [white arrow in Fig.~\ref{Fig:Map}(a)] but not at $\mathbf{Q}=(0.3, 0.3, 0)$ [Fig.~\ref{Fig:Map}(c)].
The $L$ modulated intensity of the spin resonance is observed only at $\hbar\omega$ = $2\Delta$ [Fig.~\ref{Fig:Map}(e)].
The intensity modulation along $L$ of the low-energy IC magnetic fluctuations indicates the presence of a weakly three-dimensional superconducting gaps.
It is worth mentioning that the low-energy magnetic fluctuations at $\mathbf{Q}=(0.7,0.7,L)$ is difficult to detect in this low-energy region because of the quickly decaying squared magnetic form factor at high $Q$ [$F^2_{\mathbf{Q}=(0.7,0.7,0.5)}/F^2_{\mathbf{Q}=(0.3,0.3,0.5)}\sim1/6$~\cite{Sr2RuO4_Neutron_Nagata_2004}] and the weak intensity of the IC magnetic fluctuations at low energies [Fig.~\ref{Fig:HighE}(c)].
For quantitative analysis on the spin resonance in Sr$_2$RuO$_4$, $\hbar\omega$ and $\mathbf{Q}$ dependencies of the INS intensity [$I(\hbar\omega)$ and $I(\mathbf{Q})$] are analyzed in detail below.

\begin{figure*}[t]
\includegraphics[width=17.28cm]{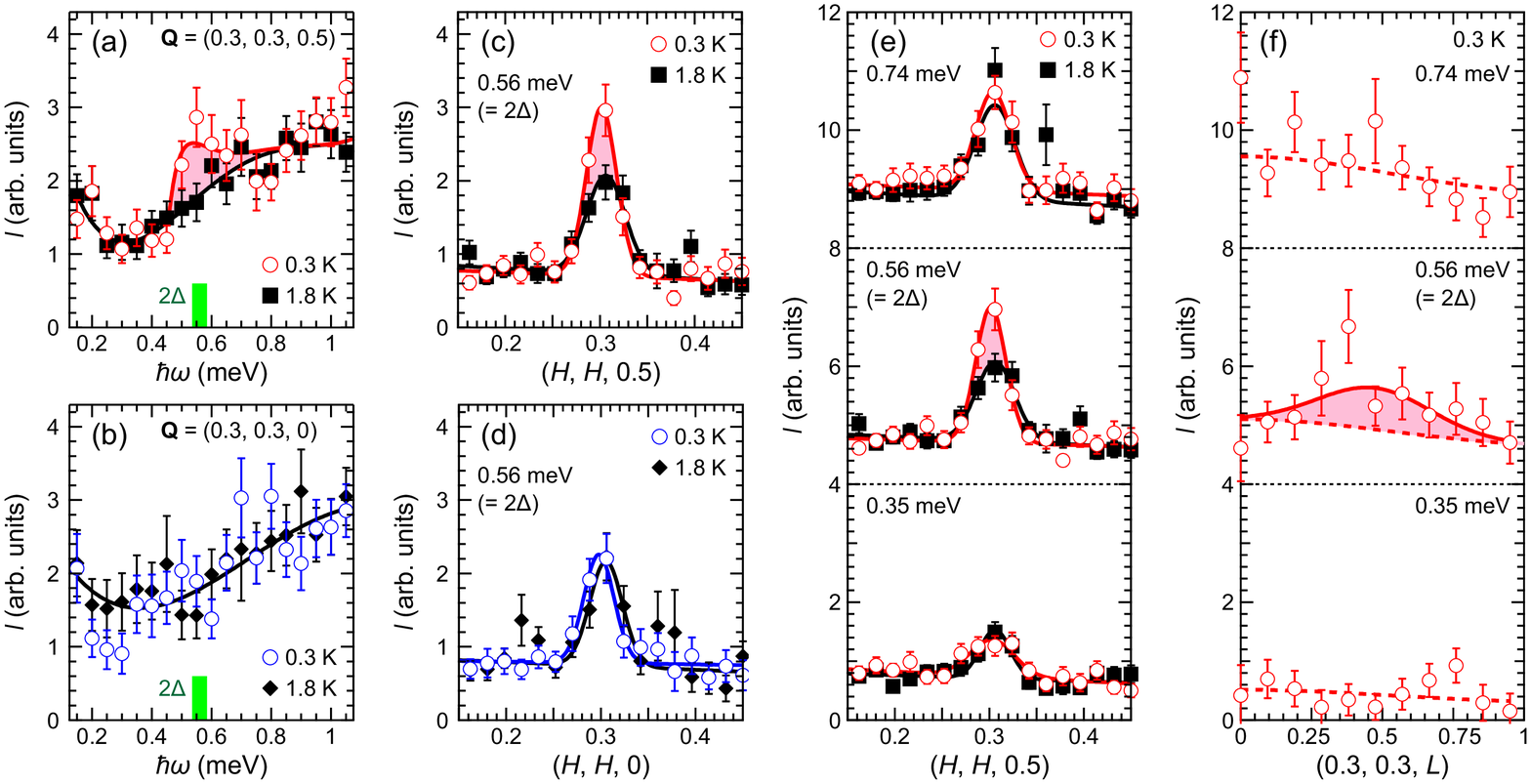}
\centering
\caption{
(Color online)
$I(\hbar\omega)$ and $I(\mathbf{Q})$ cuts of the low-energy IC magnetic fluctuations in Sr$_2$RuO$_4$.
(a, b) $I(\hbar\omega)$ cuts at (a) $\mathbf{Q}=(0.3, 0.3, 0.5)$ and (b) (0.3, 0.3, 0) below and above $T_\text{c}$.
Solid lines are the guides for the eye.
Vertical vars represent the superconducting gap $2\Delta=0.56$~meV~\cite{Sr2RuO4_Tunneling_Suderow_2009}.
(c, d) $I(\mathbf{Q})$ cuts along (c) $\mathbf{Q}=(H, H, 0.5)$ and (d) $(H, H, 0)$ at 0.3 and 1.8~K with the energy window of [0.47, 0.65]~meV.
Solid lines are the fitting results by the Gaussian function with linear background.
(e) $I(\mathbf{Q})$ cuts along $\mathbf{Q}=(H, H, 0.5)$ at 0.3 and 1.8~K with the energy windows of $[0.23, 0.47]$, $[0.47, 0.65]$, and $[0.65, 0.83]$~meV.
Solid lines are the fitting results.
(f) $I(\mathbf{Q})$ cuts along $\mathbf{Q}=(0.3, 0.3, L)$ at 0.3~K with the energy windows of [0.23, 0.47], [0.47, 0.65], and [0.65, 0.83]~meV.
Background estimated by $I(\mathbf{Q})$ along $\mathbf{Q}=(0.525, 0.525, L)$ at 0.3~K is subtracted from each spectrum.
Dashed lines are the squared magnetic form factor of Sr$_2$RuO$_4$~\cite{Sr2RuO4_Neutron_Nagata_2004} and therefore represent the components of the nearly two-dimensional IC magnetic fluctuations.
The solid line is the guides for the eye.
Shaded areas in panels (a, c, e, f) indicate the spin resonance.
}\label{Fig:Cut}
\end{figure*}

$I(\hbar\omega)$ cuts at $\mathbf{Q}=(0.3, 0.3, 0.5)$ and $(0.3, 0.3, 0)$ are plotted in Figs.~\ref{Fig:Cut}(a) and \ref{Fig:Cut}(b), respectively.
Below $T_\text{c}$, a clear increment in intensity at $\hbar\omega\sim0.56$~meV can be seen in the $I(\hbar\omega)$ cut at $(0.3, 0.3, 0.5)$ [shaded area in Fig.~\ref{Fig:Cut}(a)].
Meanwhile, such enhancement at $\hbar\omega\sim0.56$~meV below $T_\text{c}$ is not observed in the $I(\hbar\omega)$ cut at $(0.3, 0.3, 0)$ [Fig.~\ref{Fig:Cut}(b)], consistent with the previous INS study~\cite{Sr2RuO4_Neutron_Kunkemoller_2017}.
$I(\mathbf{Q})$ cuts at $\hbar\omega=0.56$~meV along $\mathbf{Q}=(H, H, 0.5)$ and $(H, H, 0)$ also show the same trend [Figs.~\ref{Fig:Cut}(c) and \ref{Fig:Cut}(d)].
The IC fluctuations at $\hbar\omega=0.56$~meV along $\mathbf{Q}=(H, H, 0.5)$ are enhanced below $T_\text{c}$ [shaded area in Fig.~\ref{Fig:Cut}(c)], while the IC fluctuations at $\hbar\omega=0.56$~meV along $\mathbf{Q}=(H, H, 0)$ do not change in intensity across $T_\text{c}$ [Fig.~\ref{Fig:Cut}(d)].
Figure~\ref{Fig:Cut}(e) shows the $I(\mathbf{Q})$ cuts along $\mathbf{Q}=(H, H, 0.5)$ for three different energies, $\hbar\omega=0.35$, 0.56, and 0.74~meV.
The $I(\mathbf{Q})$ cuts exhibit peaks centered at $\mathbf{Q}_\text{IC}=(0.3, 0.3, 0.5)$ owing to the IC magnetic fluctuations in all the energy windows at both temperatures.
Interestingly, only the INS intensity at $\hbar\omega=0.56$~meV shows sizable enhancement at $\mathbf{Q}_\text{IC}$ in the superconducting state compared to the normal state [shaded area in Fig.~\ref{Fig:Cut}(e)].
We emphasize that the energy, $\hbar\omega=0.56$~meV, corresponds well to the superconducting gap $2\Delta$~\cite{Sr2RuO4_Tunneling_Suderow_2009} [vertical bar in Fig.~\ref{Fig:Cut}(a)], suggesting that the observed enhancement, localized in both $\mathbf{Q}_\text{res}=(0.3, 0.3, 0.5)$ and $\hbar\omega_\text{res}=0.56$~meV, is the spin resonance [see also white arrow in Fig.~\ref{Fig:Map}(a)].
The spin gap of the IC magnetic fluctuations at $\mathbf{Q}_\text{IC}=(0.3, 0.3, 0.5)$ is smaller than 0.2~meV [Fig.~\ref{Fig:Cut}(a)], which makes the spin resonance less pronounced.
$I(\mathbf{Q})$ cuts at 0.74 meV in Fig.~\ref{Fig:Cut}(e) also shows a tiny increment at 0.3~K.
A possible scenario is that the reported superconducting gap value~\cite{Sr2RuO4_Tunneling_Suderow_2009} is band averaged and the superconducting gaps regarding the $\alpha$ and $\beta$ bands are slightly larger than $2\Delta=0.56$~meV.

$L$ dependence of the low-energy IC magnetic fluctuations can provide us a compelling evidence to clarify the nodal nature of the superconducting gaps~\cite{Sr2RuO4_Theory_Machida2}.
Figures~\ref{Fig:Map}(d)--\ref{Fig:Map}(f) illustrate contour maps of the INS intensities in the $(HHL)$ plane at 0.3~K with the energies 0.35, 0.56, and 0.74~meV, and $I(\mathbf{Q})$ cuts along $\mathbf{Q}=(0.3, 0.3, L)$ at 0.3~K with the corresponding energies are also plotted in Fig.~\ref{Fig:Cut}(f).
In contrast to the monotonically decreasing intensities of the IC magnetic fluctuations at 0.35 and 0.74~meV following the squared magnetic form factor of Sr$_2$RuO$_4$~\cite{Sr2RuO4_Neutron_Nagata_2004} [the dashed lines in Fig.~\ref{Fig:Cut}(f)], the $I(\mathbf{Q})$ cut at the spin resonance energy (0.56~meV) shows a maximum at $L\sim0.5$ besides the monotonically decreasing component of the IC magnetic fluctuations [the shaded area and the dashed line in Fig.~\ref{Fig:Cut}(f)].
The $L$ modulated spin resonance represents the gap symmetry as the feedback effect from the superconducting gaps.
Also, the $L$ modulated intensity explains the absence of the spin resonance at $\mathbf{Q}=(0.3, 0.3, 0)$ [Figs.~\ref{Fig:Map}(c), \ref{Fig:Cut}(b), and \ref{Fig:Cut}(d)].
To theoretically elucidate the origin of the $L$ modulated intensity of the spin resonance in Sr$_2$RuO$_4$, RPA calculations assuming the horizontal line nodes at the superconducting gaps were further performed.

To construct a realistic model, we perform density functional theory (DFT) calculations using the Wien2k package~\cite{w2k}.
We obtain an effective 3-orbital model considering the Ru $d_{xz}$, $d_{yz}$, $d_{xy}$-orbitals using the maximum localized Wannier functions~\cite{Wannier90}.
The generalized gradient approximation (GGA) exchange-correlation functional~\cite{PBE} is adopted with the cut-off energy $RK_\textrm{max}=7$ and 512 $k$-point mesh.
We renormalize the bandwidth considering the effective mass $m^*=3.5$, and the resulting renormalized bandwidth is $W\sim 1.05$~eV. 
We consider the following gap function with horizontal line nodes:
\begin{equation}
{\Delta}(\mathbf{k})=\Delta_0\cos ck_z\label{gap}
\end{equation}
within the standard BCS framework.
We take the gap amplitude $\Delta_0=4.8\times 10^{-3}W$.
In the body center tetragonal system, the period along to the $k_z$ axis is $4\pi/c$, and thus we take $k_z$ as $0\leq k_z<4\pi/c$.
We obtain the dynamical spin susceptibility $\chi^\textrm{total}_s(\mathbf{q},\omega)$ applying RPA as
\begin{equation}
\chi^\textrm{total}_s(\mathbf{q},\omega)=\sum_{l,m} \chi_s^{l,l,m,m}(\mathbf{q},\omega)
\end{equation}
\begin{equation}
\hat{\chi}_s(\mathbf{q},\omega)=\hat{\chi}_0(\mathbf{q},\omega)[\hat{I}-\hat{S}\hat{\chi}_0(\mathbf{q},\omega)]^{-1}
\end{equation}
\begin{equation}
\hat{\chi}_0(\mathbf{q},\omega)=\hat{\chi}_{0,G}(\mathbf{q},\omega)+\hat{\chi}_{0,F}(\mathbf{q},\omega)
\end{equation}
\begin{eqnarray}
\chi^{^{l_1,l_2,l_3,l_4}_{\sigma_1,\sigma_2,\sigma_3,\sigma_4}}_{0,G(F)}(\mathbf{q},\omega)&=&\sum_{k}\sum_{n,m}\frac{f(E^n_{\mathbf{k}+\mathbf{q}})-f(E^m_{\mathbf{k}})}{\omega+i\delta-E^n_{\mathbf{k}+\mathbf{q}}+E^m_{\mathbf{k}}}\label{eqchi0}\\
& &\times U_{l_1,\sigma_1,n}(\mathbf{k}+\mathbf{q})U_{l_4,\sigma_4,m}(\mathbf{k})\nonumber\\
& &\times U^{\dagger}_{m,l_2,\sigma_2}(\mathbf{k})U^{\dagger}_{n,l_3,\sigma_3}(\mathbf{k}+\mathbf{q})\nonumber
\end{eqnarray}
where $l_1\sim l_4$ and $\sigma_1\sim\sigma_4$ are the orbital ($d_{xz}, d_{yz}, d_{xy}$) and spin ($\uparrow$ and $\downarrow$) indices.
$\hat{\chi}_{0,G(F)}$ denotes the normal (anomalous) part of the irreducible bare susceptibility $\hat{\chi}_{0}$ at $\sigma_1=\sigma_2=\sigma_3=\sigma_4$ ($\sigma_1=\sigma_2\neq\sigma_3=\sigma_4$).
$E^n_{\mathbf{k}}$ and $f(E^n_\mathbf{k})$ are the eigenvalue and Fermi distribution function of Bogoliubov quasi-particles.
$\hat{S}$ is the interaction vertex matrix~\cite{Suzuki}.
We consider the on-site intra- and inter-orbital Coulomb interactions $U_l$ and $U'$ as $U_{dxz/dyz}=0.21W$, $U_{d_{xy}}=0.7U_{dxz/dyz}$, $U'=3U_{dxz/dyz}/4$.
The Hund's coupling and pair hopping $J=J'=U_{dxz/dyz}/8$.
We take the temperature $T=1.9\times 10^{-3}W$, the smearing factor $\delta=1.9\times 10^{-3}W$ and $1024\times 1024\times 32$~$k$-mesh.
The resulting characteristic spectral features of the dynamical spin susceptibly are not much influenced by the detailed parameter values mentioned above.

\begin{figure}[t]
\includegraphics[width=8.4cm]{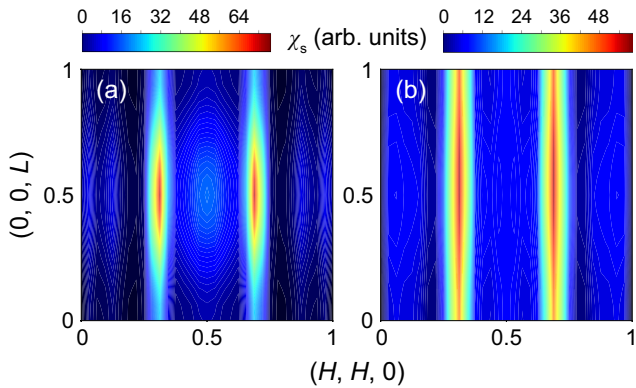}
\centering
\caption{
(Color online)
Calculated density maps of the dynamical spin susceptibility assuming the horizontal line nodes in Sr$_2$RuO$_4$.
(a) $\hbar\omega=2\Delta_0$. (b) $\hbar\omega=4\Delta_0$.
}\label{Fig:Calculation}
\end{figure}

Figures~\ref{Fig:Calculation}(a) and \ref{Fig:Calculation}(b) illustrate the calculated dynamical spin susceptibilities for energies, $\hbar\omega=2\Delta_0$ and $4\Delta_0$.
We note that our calculations do not include the squared magnetic form factor of Sr$_2$RuO$_4$ for clear visualization of the $L$ dependence.
For $\hbar\omega=2\Delta_0$, the dynamical spin susceptibility shows the maximum at $\mathbf{Q}=(1/3, 1/3, 1/2)$ and (2/3, 2/3, 1/2) as illustrated in Fig.~\ref{Fig:Calculation}(a).
The superconducting gaps with the horizontal line nodes described in Eq.~(\ref{gap}) give such the feature along $L$, which is indeed observed in our INS measurements [Fig.~\ref{Fig:Map}(e)].
We address that the observed $L$ modulated intensity cannot be explained by the superconducting gap with the vertical line nodes since the superconducting gap changes its sign within the $k_z$ plane~\cite{Sr2RuO4_Theory_Machida2}.
On the other hand, no pronounced $L$ modulation at higher energy transfers is observed either in the experiment [Fig.~\ref{Fig:Map}(f)], except the monotonically decreasing intensities due to the magnetic form factor, or in the calculation [Fig.~\ref{Fig:Calculation}(b)].
Since the energy window is too high compared to the superconducting gaps in Sr$_2$RuO$_4$~\cite{Sr2RuO4_Tunneling_Suderow_2009}, the effect of the superconducting gaps smears out, and only the quasi-two-dimensional feature of the IC magnetic fluctuations show up in the higher energy windows ($\hbar\omega\gg2\Delta$).

The observed spin resonance at $\mathbf{Q}_\text{res}=(0.3, 0.3, 0.5)$ and $\hbar\omega_\text{res}$ = $2\Delta$ indicates that the quasi-one-dimensional $\alpha$ and $\beta$ sheets are active bands for the bulk superconductivity of Sr$_2$RuO$_4$.
Our RPA calculations reveal that the observed $L$ modulated intensity of the spin resonance originates from the horizonal line nodes at the superconducting gaps, in agreement with the field-angle-dependent specific heat capacity measurements~\cite{Sr2RuO4_Magnetization_Kittaka_2014}.
We believe that our results give a strong constraint on the superconducting pairing symmetry of Sr$_2$RuO$_4$.

Let us now discuss the possible pairing symmetry of Sr$_2$RuO$_4$.
Among the pairing symmetries with the horizontal line nodes proposed by the recent uniaxial pressure~\cite{Sr2RuO4_Pressure_Hicks_2014,Sr2RuO4_Pressure_Steppke_2017,Sr2RuO4_Pressure_Watson_2018,Sr2RuO4_Pressure_Barber_2018,Sr2RuO4_Pressure_Luo_2019,Sr2RuO4_Pressure_Sunko_2019,Sr2RuO4_Pressure_Li_2019} and NMR works~\cite{Sr2RuO4_newNMR_Pustogow_2019,Sr2RuO4_newNMR_Ishida_2019}, the chiral $d$-wave $(k_x + ik_y)k_z$~\cite{Theory_Zutic} is compatible with the time-reversal symmetry breaking~\cite{Sr2RuO4_muSR_Luke_1998,Sr2RuO4_Kerr_Xia_2006}.
However, its wave function is odd in the $k_z$ plane giving rise to the spin resonance at $\mathbf{Q}=(0.3, 0.3, 0)$, and thus such pairing symmetry can be excluded by our INS data.
Recently, the Josephson effects measurements suggested the time-reversal invariant superconductivity in Sr$_2$RuO$_4$~\cite{Sr2RuO4_newJosephson_Kashiwaya_2019}.
Combined with the time-reversal invariant superconductivity and the horizontal line nodes, $d$-wave $d_{3k_z^2-1}$~\cite{Sr2RuO4_Theory_Machida2} is proposed for the candidate for the pairing symmetry of Sr$_2$RuO$_4$.
The single component order parameter is also in line with the absence of a split transition in the presence of uniaxial strain and $T_\text{c}$ cusp in the limit of zero strain~\cite{Sr2RuO4_Pressure_Hicks_2014,Sr2RuO4_Pressure_Steppke_2017,Sr2RuO4_Pressure_Watson_2018}.
We note that the horizontal line nodes assuming $d_{3k_z^2-1}$ is not symmetry protected.
However, the horizontal line nodes can be realized even if the $d$-wave $d_{3k_z^2-1}$ belonging to the $A_{1g}$ representation in the point group $D_{4h}$ slightly mixes with the $s$ wave.
Here, we emphasize that the spin resonance at $L=0.5$ is more sensitive to the gap modulation along $k_z$ than the $k_z$ position of horizontal line nodes.
The candidate of the origin of the horizontal nodes is a competition of pairing mechanisms.
Since spin fluctuations are highly two-dimensional, it would be difficult to arise three-dimensional gap structure only with pure spin fluctuation mechanism.
The crystal structure of Sr$_2$RuO$_4$ is body-centered tetragonal.
Therefore, it seems that the electron-phonon~\cite{Phonon} mediated mechanism would have three-dimensionality.
The observed weakly three-dimensional spin resonance at $L = 0.5$ may come from a competition between spin fluctuations and electron-phonon coupling.

In summary, we investigated the low-energy IC magnetic fluctuations in Sr$_2$RuO$_4$ below and above $T_\text{c}$ with detailed analysis on the $L$ dependence.
Below $T_\text{c}$, the spin resonance appears at $\mathbf{Q}_\text{res}=(0.3, 0.3, 0.5)$ and $\hbar\omega_\text{res}=0.56$~meV.
The spin resonance shows the $L$ modulated intensity with a maximum at $L\sim0.5$.
The $L$ modulated intensity of the spin resonance and our RPA calculations indicate that the superconducting gaps regarding the $\alpha$ and $\beta$ sheets have horizontal line nodes.

The experiments at AMATERAS were conducted under the J-PARC MLF user program with the proposal numbers 2018A0060 and 2018AU1402.
The present work was supported by JSPS KAKENHI Grant Numbers JP17K05553 and JP17K14349, and the Cooperative Research Program of ``Network Joint Research Center for Materials and Devices'' (20181072).

\end{document}